\documentclass[12pt]{book}

\usepackage[dvips]{graphicx,color}
\usepackage{makeidx,tsukuba}

\makeauthorindex
\makeindex

\begin{document}

\BookTitle{\itshape The 28th International Cosmic Ray Conference}
\CopyRight{\copyright 2003 by Universal Academy Press, Inc.}
\pagenumbering{arabic}

\chapter{
The Trigger System of the ARGO-YBJ detector}

\author{%
%
%
S. Mastroianni,$^1$ A. Aloisio,$^2$ S. Catalanotti,$^1$ S. Cavaliere,$^1$ P. Bernardini,$^3$ P. Creti,$^3$ I. De Mitri,$^3$ G. Marsella,$^3$ M. Panareo,$^3$ A. Surdo,$^3$  for the ARGO-YBJ collaboration [1]     \\
{\it (1) Dip. di Scienze Fisiche Universit\'a di Napoli Federico II and INFN sez. di Napoli, Napoli, Italy \\
(2) Universit\'a del Sannio, Benevento and INFN sez. di Napoli, Napoli, Italy \\
(3) Dip. di Fisica  Universit\'a di Lecce and INFN sez. di Lecce, Lecce, Italy} \\
}

\section*{Abstract}

The ARGO-YBJ experiment has been designed to detect air shower events over a large size scale and with an energy
threshold of a few hundreds GeV. The building blocks of the ARGO-YBJ detector are single-gap Resistive Plate
Counters (RPCs). The trigger logic selects the events on the basis of their hit multiplicity. Inclusive triggers
as well as dedicated triggers for specific physics channels or calibration purposes have been developed. This paper describes the
architecture and the main features of the trigger system.

\section{Introduction}

The ARGO-YBJ experiment [1] is devoted to a wide range of fundamental issues in
cosmic rays and astroparticle physics, including in particular $\gamma$-ray astronomy
and $\gamma$-ray bursts physics at energies $\geq$ 100 GeV. The detector consists of an array of single layer RPCs
that covers $74 \times 78~ m^{2}$, surrounded by a partially instrumented
guard ring. It is organized in modules of 12 chambers
called CLUSTER. Each chamber is made of 80 strips and the entire detector
comprises 130 CLUSTERs in the central carpet. The 960 strips in each CLUSTER
are processed by the CLUSTER logic (fig. 1). For time measurement and trigger
purposes, adjacent strips are logically OR-ed together in groups of 8 defining
a logic unit called PAD of $56 \times 62~ cm ^{2}$ (15600 in total). The PAD signals
are stretched to 150 ns to guarantee that the particles of the same shower are
in coincidence. The CLUSTER logic outputs a 6-bit Low Multiplicity weighted bus
(when $\geq 1$, $\geq 2$, $\geq 3$, $\geq 4$, $\geq 5$, $\geq 6$ PADs are fired) and a 4-bit
High Multiplicity weighted bus (when $\geq 7$, $\geq 16$, $\geq 32$, $\geq 64$  PADs are fired).
The CLUSTER logic is housed in a location nearby the
CLUSTER, called Local Station (LS).

The PADs provide a detailed space-time picture of the shower front and they are
the basic inputs to the trigger logic, which validates an event on the basis of
specific fired PADs density criteria.

The ARGO-YBJ experiment has been designed to
study a wide range of fundamental issues in cosmic ray physics and $\gamma$-ray
astronomy. The energy experimental range is very large, bridging the GeV and TeV
energy regions of the showers, and moreover the particle density distribution
changes with the distance from the shower axis.
Showers with very low energy, in the range of a few hundreds GeV, are expected
to fire less than 100 PADs spread on the entire carpet (fig. 2). At very high energy, in
the range of tenths of TeV and beyond, the showers present a specific spatial
distribution, characterized by a core structure. The trigger logic should be
able to select events exceeding a programmable threshold of PADs, fired by the
same shower. At low threshold, such algorithm allows us to trigger showers with
arbitrary hit density. Raising the thresholds,
showers at higher energy can be triggered. A local density trigger can be obtained
by applying the same logic
to each detector partition made of four adjacent clusters (SUPERCLUSTER). This
local density concept can be extended to an arbitrary CLUSTER pattern, requiring
that a programmable number of clusters exceeds a specific threshold setting to produce a trigger.
These three algorithms cover the intended physics range of interest for the
ARGO-YBJ experiment, featuring a trigger scheme based on both event density
and hit multiplicity.

\begin{figure}[t]
  \begin{center}
    \includegraphics{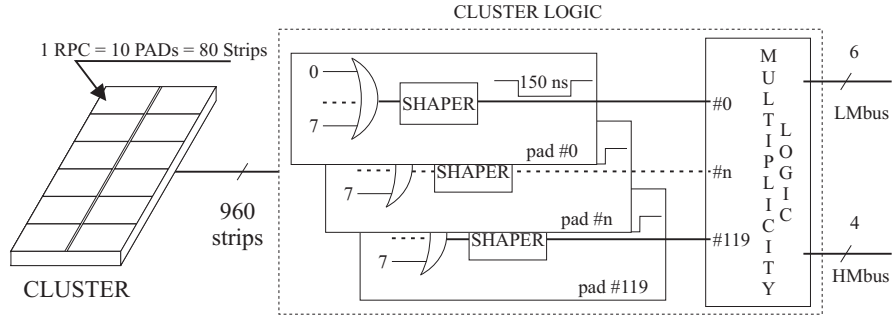}
  \end{center}
  \vspace{-0.5pc}
  \caption{The Cluster Logic}
\end{figure}

\section{The Trigger Architecture}

The three different trigger logics are based on the Low Multiplicity (LM) and
High Multiplicity (HM) busses driven by  the LSs. In order to preserve the
original data timing, all the hardware is implemented as synchronous pipelines
with a common system clock at 33 MHz, generated and distributed by a specific
low-jitter balanced clock tree.

The first trigger logic generates a signal when a programmed threshold of fired PADs is
reached or exceeded on the entire detector (LM Trigger).  The  LM Trigger uses the LMbus from
each LS and processes from the central carpet a total of $130 \times 6$ bit inputs. In order to handle a data
word that wide, the logic adopts a three-level concentration scheme. The Level-1
 is based on the LOWM boards. This module synchronizes the LMbusses from 12 LSs
to the system clock,  stretches the bit width to 8 clock cycles ($\sim$ 243 ns)  in order
to garantee the  coincidence across the 12 CLUSTERs in the worst case hit timing
and outputs on a 7-bit binary encoded bus the total multiplicity, updated every clock cycle.
The central carpet is split in two regions as shown in Fig. 3. Each region comprises 65
CLUSTERs processed by 6 LOWM boards.
In the same fashion, two Level-2 boards called
$\Sigma$LOW stretch the Level-1 7-bit outputs to 12 clock cycles ($\sim$ 360 ns) and output on a 10-bit binary encoded bus
the total multiplicity for each half carpet. At the end, a single Level-3 board TLOW shapes the Level-2 10-bits outputs
to about 400 ns and encodes the total 11-bit multiplicity for the entire carpet.
This board generates the low multiplicity trigger when the hit total number exceedes the programmed threshold on the carpet.
Each LOWM splits a 12 CLUSTER region in 3 SUPERCLUSTERs. A specific threshold can be assigned to each SUPERCLUSTER and the module generates a
 Fast Trigger if at least a SUPERCLUSTER hit number exceedes its correspondent threshold.
\begin{figure}[t]
  \begin{center}
    \includegraphics[height=12.5pc]{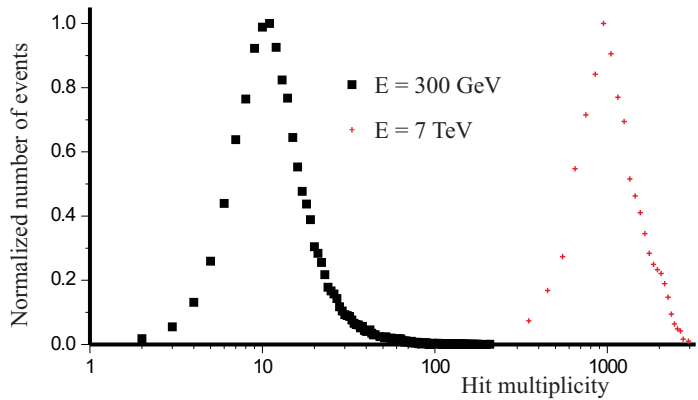}

  \end{center}
  \vspace{-0.5pc}
  \caption{Number of events versus Hit multiplicity}
\end{figure}
\indent
The third trigger logic produces an output when a programmable threshold of CLUSTERs with
a specific multiplicity in the range coded by the HMbus is excedeed (HM Trigger). For instance,
this logic allow us to trigger on events where at least N CLUSTERs present a hit multiplicity $\geq 7$.
The HM Trigger is split in four sections, each handling  in parallel a specific bit from each LS's 4-bit
HMbus: the HM 7, 16, 32, 64 Triggers. Each section processes from the central carpet a 130-bit input
and it is implemented in two levels as shown in fig. 4. Each HIGHM board synchronizes the i-th bits of
the HMbusses from 65 LSs to the system clock. Also, it stretches the bit width to 12 clock cycle ($\sim$ 360 ns) in order
to count the CLUSTERs in  coincidence across half carpet with the pertaining hit multiplicity.
The two 7-bit binary encoded busses driven by the HIGHM boards are merged into the THIGH board
that calculates the total number of CLUSTERs across the entire carpet.
The LM Trigger, the Fast Trigger and the four HM Triggers are handled by a TRBOX board which can mask
a specific trigger source and OR them together. If the Data Acquisition System does not assert a Veto signal,
the trigger proposal becomes a trigger signal. Apart from the system
initialization, the trigger system activity is fully self governing.

 \begin{figure}[t]
\vfill \begin{minipage}[t]{.47\linewidth}
 \begin{center}
   \includegraphics[height=13.5pc]{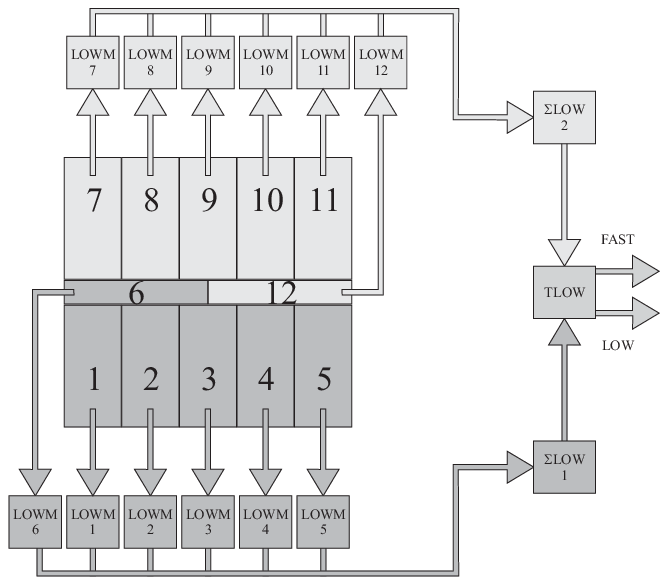}
 \end{center}
 \vspace{-0.5pc}
   \caption{The Low Multiplicity Trigger logic.}
  \label{llf1_llf2}
\end{minipage}\hfill
\hspace{-0.5cm}
\begin{minipage}[t]{.47\linewidth}
 \begin{center}
   \includegraphics[height=13.5pc]{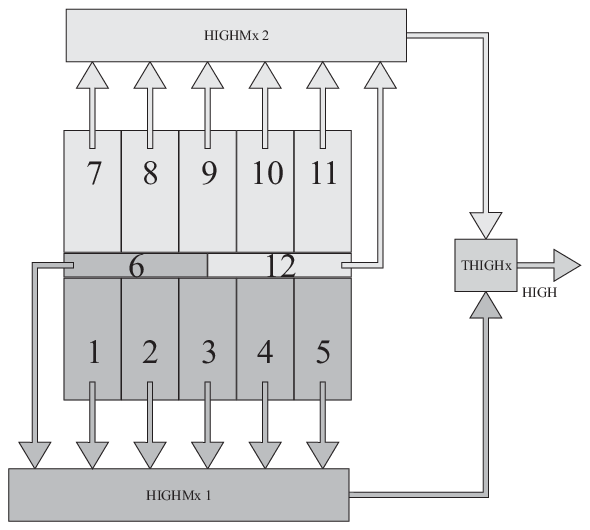}
 \end{center}
 \vspace{-0.5pc}
   \caption{The High Multiplicity Trigger logic.}
  \label{finout}
\end{minipage}\hfill
\end{figure}

\section{The trigger system boards}

All the trigger boards are VME double height slaves with A32/D08(O), D16, D32,
and D32:BLT data transfer capabilities. Differential data
in P(ositive) ECL standard assures a superior performance in terms of sustainable data
rates and a cable driving capability using cheap twisted-pair media. Through
the VME bus the main trigger features and operating modes can be programmed and monitored.

\section{Conclusion}

In the realization of these modules we started from the beginning using VHDL and
synthesis tools. The use of the FPGAs give to us enough flexibility giving an easy
way to upgrade existing boards with new features when necessary.

\section{References}


\re
1.\ Surdo A. et al.\ 2003, in this proceedings.

\end{document}